\documentclass[useAMS,usenatbib]{mn2e}
\usepackage{amsmath,amssymb}
\usepackage{aas_macros}
\usepackage{graphicx}
\usepackage{times}
\def\ifnote{\iffalse}
\def\dd{ {\rm d} }

\title[Testing EEP with polarized GRBs]{Testing the Einstein's equivalence principle with polarized gamma-ray bursts}
\author[C.Yang \it et.\,al.]{Chao Yang$^{1}$, Yuan-Chuan Zou$^{1}$\thanks{Email: zouyc@hust.edu.cn}, Yue-Yang Zhang$^{1}$, Bin Liao$^{1}$, Wei-Hua Lei$^{1}$
\\
$^{1}${School of Physics,
Huazhong University of Science and Technology,  Wuhan 430074, China}
}

\begin{document}
\date{\today}
\maketitle
\label{firstpage}

\begin{abstract}
The Einstein's equivalence principle can be tested by using parameterized post-Newtonian parameters, of which the parameter $\gamma$ has been constrained by comparing the arrival times of photons with different energies. It has been constrained by a variety of  astronomical transient events, such as gamma-ray bursts (GRBs), fast radio bursts as well as pulses of pulsars, with the most stringent constraint of $\Delta \gamma \lesssim 10^{-15}$. In this letter, we consider the arrival times of lights with different circular polarization. For a linearly polarized light, it is the combination of two circularly polarized lights. If the arrival time difference between the two circularly polarized lights is too large, their combination may lose the linear polarization. We constrain the value of $\Delta \gamma_{\rm p} < 1.6 \times 10^{-27}$ by the measurement of the polarization of GRB 110721A, which is the most stringent constraint ever achieved.
\end{abstract}


\section{Introduction}\label{sec:intr}
The Einstein's  equivalence principle (EEP) is an important foundation of metric theories of gravity. EEP can be tested through parameterized post-Newtonian (PPN) parameters, of which the parameter $\gamma$ indicates how much space curvature is produced by unit rest mass \citep{2014LRR....17....4W}. One statement of EEP is that any uncharged test body traveling in empty space follows a trajectory independent of its internal structure and composition. Therefore, if EEP is satisfied, the parameter $\gamma$ will be the same for different particles, and two objects traveling through the same distance will arrive at the same time.

The arrival time delays between photons with different energies emitted from astronomical sources have been widely used to test EEP by constraining the parameter $\gamma$ discrepancy between different photons. The time delay values are usually obtained in the following two methods. First, the light curves of photon fluxes with different energies have similar features for a source, e.g. a gamma-ray burst \citep[GRB,][]{2015ApJ...810..121G, 2016ApJ...821L...2N}, a TeV Blazer \citep{2016ApJ...818L...2W} and pulses of a pulsar \citep{2016arXiv161200717Z}. The time delay is measured by cross correlation of two light curves. Second, when a burst event lasts a very short time, e.g. a short GRB \citep{2016MNRAS.460.2282S}, a fast radio burst \citep[FRB,][]{2015PhRvL.115z1101W,2016ApJ...820L..31T}, or a giant pulse of a pulsar \citep{2016PhRvD..94j1501Y}, the burst duration time is applied as the time delay between the highest and lowest energies within the bandpass of the observing telescope.

By far, the most stringent constraint obtained by the first method and second mothod are $\gamma_{\rm GeV}-\gamma_{\rm MeV}<4\times10^{-11}(3\sigma)$, or $ \gamma_{\rm GeV}-\gamma_{\rm MeV} < 2.3\times10^{-12} (2\sigma)$, for GRB 090510 with a high redshift of $z=0.903\pm0.003$ \citep{2016ApJ...821L...2N}, and $\gamma(\rm 10.35~GHz)-\gamma(\rm 8.15~GHz)< (0.6-1.8)\times10^{-15}$ , using a 0.4-nanosecond giant burst of the Crab pulsar \citep{2016PhRvD..94j1501Y}, respectively.

In this letter, we report a measurement of the time delay between lights with different circular polarization. This method has been applied to constrain the Lorentz invariance violation (LIV)  \citep{2007MNRAS.376.1857F, 2012PhRvL.109x1104T}. Linearly polarized light is a superposition of two opposite circularly polarized lights. Its polarization angle  rotates $\Delta \phi$ if the two beams of light arrive with a time difference. We propose a method to constrain the parameter $\gamma$, by using linearly polarized light from GRBs. In the rest of this letter, we will describe our method in $\S2$, and apply it to GRBs in $\S3$.  The conclusion and discussion will be  presented in $\S4$.

\section{METHODOLOGY}\label{sec:method}
In PPN approximation, Shapiro time delay in a gravitational potential $U({\bf r})$ is given by \citep{1964PhRvL..13..789S,1988PhRvL..60..176K,1988PhRvL..60..173L}
\begin{equation}\label{1}
\delta t_{\rm gra} = -\frac{1+\gamma}{c^{3}}\int_{\bf r_{\rm e}}^{\bf r_{\rm o}}U(\bf r) d \bf r,
\end{equation}
where $\bf r_{\rm e}$ and $\bf r_{\rm o}$ are the locations of the source and the observer, respectively.
We consider a linearly polarized light, which is composed by two circularly polarized beams (labeled with `r' and `l'). If the two beams pass through the same gravitational potential with different Shapiro time delays because of different $\gamma$ ($\gamma_{\rm l} \neq \gamma_{\rm r}$), the time lag of these two beams is then
\begin{equation}\label{2}
\Delta t_{\rm gra} = \mid\frac{\Delta \gamma_{\rm p}}{c^{3}}\int_{\bf r_{\rm e}}^{\bf r_{\rm o}}U(\bf r) d \bf r\mid,
\end{equation}
where $\Delta \gamma_{\rm p} \equiv \gamma_{\rm l}-\gamma_{\rm r}$ is the different of $\gamma$ for the left polarized light and the right polarized light. The time lag results in the rotation of the linear polarization angle as
\begin{equation}\label{3}
\Delta \phi = \Delta t_{\rm gra}  \frac{2 \pi c}{\lambda}.
\end{equation}

The exact value of $\Delta \phi $ is unknown because the initial angle of the polarized light is not available. Yet, we can set an upper limit for  $\Delta \phi $ so that it cannot exceed more than $2 \pi$, otherwise, the light will become unpolarized as the path difference goes beyond the coherence length. If one observes some object with linear polarization, it indicates $\Delta \phi < 2 \pi$ \footnote{A constraint of time lag by the polarization has also been used in the LIV constraint \citep{2007MNRAS.376.1857F,2012PhRvL.109x1104T}. Here we suppose that the LIV effect does not work simultaneously to accidentally cancel the effect from the EEP violation.}. This puts an upper limit on the  parameter $\gamma$
\begin{equation}\label{5}
 \Delta\gamma_{\rm p} < \frac{ c^2}{|\int_{\bf r_{\rm e}}^{\bf r_{\rm o}}U(\bf r)d\bf r|} \lambda.
\end{equation}
It can be seen that, with the shorter wavelength, the constraint is more stringent.

\section{Application to GRBs}
As shown above, the most stringent constraint can be obtained from the highest energy band of polarization observations. There are several GRBs  with reported polarization at  $\gamma$-ray band \citep[see][for a review]{2017NewAR..76....1M}, such as GRB 930131 with linear polarization degree $\Pi = (35-100 \%)$ \citep{2005A&A...439..245W}, GRB 041219a with $\Pi=96\% \pm 40\%$ \citep{2007A&A...466..895M}, GRB 110301A with $\Pi=70\% \pm 22\%$ \citep{2012ApJ...758L...1Y} and GRB 110721A with $\Pi = 84^{+16}_{-28}\%$   \citep{2012ApJ...758L...1Y}. Here GRB 110721A is selected as it has a high linear polarization with low uncertainty. The redshift of the source has been measured \citep{2011GCN..12193...1B}, i.e. $z=0.382$, corresponding to a comoving distance of $d=4.6\times10^{27}$ cm. \footnote{Though the measured redshift may not be from the host galaxy of the GRB 110721A because  the X-ray and optical counterparts lie outside the the inter-planetary network (IPN) error box \citep{2011GCN..12195...1H}, and the upper limit $z=3.512$ for the GRB comes from the Ly-$\alpha$ absorption  \citep{2011GCN..12193...1B}. The actual distance does not change the result much, as one can see from Equation (\ref{6}). We choose $z = 0.382$ for a conservative constraint on $\Delta \gamma_{\rm p}$. } 

The gravitational potential of a cosmological source is $U({\bf r}) = U_{\rm MW}({\bf r})+U_{\rm IG}({\bf r})+U_{\rm host}({\bf r})$, where $U_{\rm MW}(r)$ is the gravitational potential of the Milky Way galaxy, $U_{\rm IG}(r)$  the intergalactic background between host galaxy and the Milky Way, and $U_{\rm host}(r)$ the host galaxy of the source. Although $U_{\rm IG}({\bf r})$ and $U_{\rm host}({\bf r})$ are unknown, the contributions of these terms are significantly less than $U_{\rm MW}({\bf r})$ as discussed in  \citet{2015ApJ...810..121G}.
Adopting the Keplerian potential of the Milky Way galaxy, namely $U_{\rm MW}(r)= -\frac{GM_{\rm MW}}{r}$,
one has \citep{2015ApJ...810..121G}
\begin{equation}\label{6}
 \int_{r_{\rm e}}^{r_{\rm o}}U({\bf r}) \dd r \simeq GM_{\rm MW}\ln \frac{d}{b},
\end{equation}
where $G=6.68 \times10^{-8}$$\rm erg$ $\rm cm$ $\rm g^{-2}$ is the gravitational constant, $M_{\rm MW}\approx 6\times10^{11}M_{\rm \odot}$ is the mass of the Milky Way \citep{2011MNRAS.414.2446M}, $d$ is the proper distance between the source and the observer, and $b$ is the impact parameter of the light rays relative to the Galactic center.
The impact parameter $b$ can be estimated as \citep{2015ApJ...810..121G}
\begin{equation}
 b = r_{\rm G}\sqrt {1-\left(\sin\delta_{\rm S}\sin\delta_{\rm G}+\cos\delta_{\rm S}\cos\delta_{\rm G}\cos(\beta_{\rm S}-\beta_{\rm G})\right)^{2}},
 \label{b}
\end{equation}
where $r_{\rm G}=8.3$ $\rm kpc$ is the distance from the Sun to the Galactic
center, $\beta_{\rm S}$ and $\delta_{\rm S}$ the right ascension and
declination of the source in  equatorial coordinates, and ($\beta_{\rm G}=17^{h}45^{m}40.04^{s},\delta _{\rm G}=-29^{\circ}00^{\prime}28.1^{\prime\prime}$)  the coordinates of the Galactic center \citep{2009ApJ...692.1075G}.

GRB 110721A was first detected by {\it Fermi}-GBM at the direction ($\beta_{\rm S}=333.66^\circ$, $\delta_{\rm S}=-38.59 ^\circ$) with uncertainty of $1^\circ$ \citep{2011GCN..11771...1F, 2011GCN..12187...1T}. The polarization was measured with IKAROS/GAP at the band of (70, 300) keV  \citep{2012ApJ...758L...1Y}. To get a conservative constraint,  the lowest energy $70$ keV is considered as the corresponding  wavelength.

Taking all the values above into Equation (\ref{5}),  the constraint of the $\gamma$ discrepancy is  $\Delta \gamma_{\rm p} < 1.6 \times 10^{-27}$.

\section{Conclusion and discussion}
By taking a beam of light with two different circular polarization (left and right) as different objects, we tested the Einstein's equivalence principle with the parameter $\gamma$. Taking GRB 110721A with high linear polarization into consideration, the $\gamma$ discrepancy between two beams with different circular polarization is constrained by $\Delta \gamma_{\rm p} < 1.6 \times 10^{-27}$, which is the most stringent constraint ever achieved. This result benefits from the fact that the phase information is taken into account. For the GRB photons at 70 keV, the wavelength is $1.8 \times 10^{-9}$ cm,  corresponding to a time lag of $\Delta t \sim 6 \times 10^{-20}$ s for a phase difference of $2\pi$. This is much shorter than the shortest time difference obtained from the light curve, such as $10^{-9}$ s for the nano-shot from the Crab pulsar \citep{2016PhRvD..94j1501Y}. Therefore, the much more stringent constraint of $\Delta \gamma$ is yielded.

This method of testing of EEP can be refined by laboratory experiments. A beam of light with linear polarization can be produced and emitted from the Earth and received by a satellite  in the space. The polarization angle $\Delta \phi $ is measured as an exact value rather than an upper limit, and can be substituted into Equations (\ref{2}) and (\ref{3}). Consequently, the parameter $\Delta \gamma_{\rm p}$ is measurable in principle.

The change of the linear polarization angle is also affected by the magnetic field. It is the so-called Faraday rotation. The dependence of $\Delta \phi$ on the Faraday rotation is $\Delta \phi \propto \lambda^2$, different from its dependence on the EEP violation as $\Delta \phi \propto \lambda^{-1}$ shown in  Equation (\ref{3}).
For the astrophysical object, if there is linear polarization measured in several bands, the Faraday term can be subtracted with the fitting of $ \lambda^2$ term. 

\section*{Acknowledgments}
We thank  Biping Gong, Wei Xie, and Wei Chen for helpful discussions, Likang Zhou and  Senlin Yu for language polishing.  This work is supported by the National Basic Research Program of China (973 Program, Grant No. 2014CB845800) and by the National Natural Science Foundation of China (Grants No. U1231101  and U1431124).

\end{document}